\documentclass[twocolumn,american,aps,prb]{revtex4}
\usepackage{mathptmx}

\usepackage[T1]{fontenc}
\usepackage[latin9]{inputenc}
\usepackage{amsmath}
\usepackage{amssymb}
\usepackage{graphicx}

\makeatletter
\@ifundefined{textcolor}{}
{%
 \definecolor{BLACK}{gray}{0}
 \definecolor{WHITE}{gray}{1}
 \definecolor{RED}{rgb}{1,0,0}
 \definecolor{GREEN}{rgb}{0,1,0}
 \definecolor{BLUE}{rgb}{0,0,1}
 \definecolor{CYAN}{cmyk}{1,0,0,0}
 \definecolor{MAGENTA}{cmyk}{0,1,0,0}
 \definecolor{YELLOW}{cmyk}{0,0,1,0}
}

\usepackage{float}

\usepackage{textcomp}

\usepackage{esint}



\makeatother

\usepackage{babel}
\begin{document}

\title{\noindent {\Large Mechanism behind self-sustained oscillations in
direct current glow discharges and dusty plasmas }}

\author{\noindent \textbf{Sung Nae Cho}}

\email{sungnae.cho@samsung.com }

\affiliation{\noindent Devices R\&D Center, Samsung Advanced Institute of Technology,
Samsung Electronics Co., Ltd, Mt. 14-1 Nongseo-dong, Giheung-gu, Yongin-si,
Gyeonggi-do 446-712, Republic of Korea. }

\date{2 April 2013 }
\begin{abstract}
\noindent An alternative explanation to the mechanism behind self-sustained
oscillations of ions in direct current (DC) glow discharges is provided.
Such description is distinguished from the one provided by the fluid
models, where oscillations are attributed to the positive feedback
mechanism associated with photoionization of particles and photoemission
of electrons from the cathode. Here, oscillations arise as consequence
of interaction between an ion and the surface charges induced by it
at the bounding electrodes. Such mechanism provides an elegant explanation
to why self-sustained oscillations occur only in the negative resistance
region of the voltage-current characteristic curve in the DC glow
discharges. Furthermore, this alternative description provides an
elegant explanation to the formation of plasma fireballs in the laboratory
plasma. It has been found that oscillation frequencies increase with
ion's surface charge density, but at the rate which is significantly
slower than it does with the electric field. The presented mechanism
also describes self-sustained oscillations of ions in dusty plasmas,
which demonstrates that self-sustained oscillations in dusty plasmas
and DC glow discharges involve common physical processes. 
\end{abstract}
\maketitle

\section{Introduction}

\begin{figure}[h]
\begin{centering}
\includegraphics[width=0.9\columnwidth]{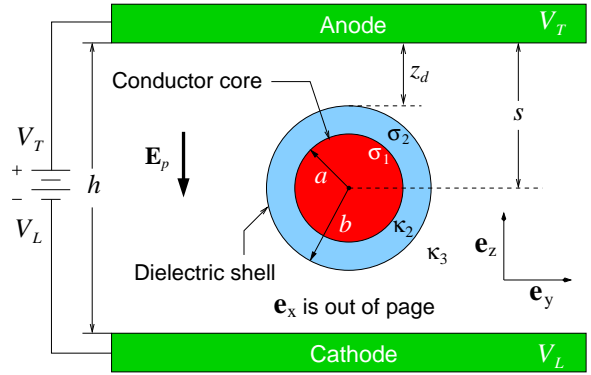}
\par\end{centering}

\caption{(Color online) Illustration of core-shell structured ion confined
between DC voltage biased plane-parallel conductors with a vacuum
gap of $h.$ $\mathbf{E}_{p}=-\mathbf{e}_{z}\left(V_{T}-V_{L}\right)/h$
is the parallel plate electric field, where ($\mathbf{e}_{x},\mathbf{e}_{y},\mathbf{e}_{z}$)
are versors along the Cartesian ($x,y,z$) axes, respectively. \label{fig:1}}
\end{figure}

When an ion is confined between the two plane-parallel electrodes
and is subject to static electric field (see Fig. \ref{fig:1}), it
begins to self-oscillate without damping. Such phenomenon is referred
to as self-sustained oscillations in direct current (DC) glow discharges;
and, it has been known for almost a century.\cite{0-Pardue-1928,0-Fox-1930,0-Fox-1931,0-Donahue-1951,0-Colli-1954,0-Pilon-1957,0-Ogawa-1959}
In literature, the phenomenon of DC glow discharge is also referred
to as the DC glow corona. The mechanism behind such oscillations is
still not fully understood.\cite{1-Sigmond-1985-EXP,1-sheat-oscillation-1996-EXP,1-Sigmond-1997-EXP,1-Petrovic-1997-EXP,1-Schoenbach-1997-EXP,1-Lozneanu-2002-EXP,1-Makarov-2006-EXP,1-fireball-UCLA-2011-EXP-m4,1-Kuschel-2011-EXP-m6}
Over the years, various theoretical models have been proposed in an
attempt to explain the phenomenon.\cite{0-Colli-1954,3-Phelps-1992-ECM,3-Hsu-2003-ECM,3-Chabert-2010-ECM,3-He-2012-ECM,2-Peres-1995-FM,2-Morrow-1997-FM,1-Sigmond-1997-EXP,2-Donko-1999-FM,2-Akishev-1999-FM,2-Allen-2007-FM}
Among the successful ones are those based on the fluid and equivalent
circuit models. 

The equivalent circuit models try to predict oscillations by representing
the system with an equivalent $\textnormal{RLC}$ circuit, where $\textnormal{R}$
is a resistor, $\textnormal{L}$ is an inductor, and $\textnormal{C}$
is a capacitor.\cite{3-Phelps-1992-ECM,3-Hsu-2003-ECM,3-Chabert-2010-ECM,3-He-2012-ECM}
Although this approach is quite useful in describing oscillation frequencies
as function of DC bias voltages, it says nothing about the mechanism
behind self-sustained oscillations. 

The fluid models approach the problem from more fundamental grounds
of the electromagnetic theory. In this approach, the Poisson equation
is solved in combination with the electron and the ion flux continuity
equations, which constitute the so called positive feedback mechanism.\cite{2-Peres-1995-FM,2-Morrow-1997-FM,2-Donko-1999-FM,2-Akishev-1999-FM,2-Allen-2007-FM}
Assertion of such feedback mechanism is crucial in the fluid models
because, without it, no oscillatory solutions can be obtained. Typical
sources of the positive feedback mechanism include the photoionization
of particles and photoemission of electrons from the cathode. Physically,
such feedback mechanism promotes oscillations by periodically reversing
the particle's charge polarity. An ion oscillating near an electrode
induces current in the same electrode, where the waveform of such
current is correlated to its velocity profile.\cite{2-Morrow-1997-FM}
Experimentally, it is this induced electrode current (or voltage)
which gets measured.\cite{0-Pilon-1957,1-Sigmond-1997-EXP} The hallmark
of the fluid models is their potential to reproduce experimental measurements
to a reasonably good accuracy. In principle, with an appropriate specification
of the positive feedback mechanism, the level of accuracy presented
by the fluid models can always be improved. The feedback mechanism
varies among different authors; and, this has been a subject area
of ongoing debate among different fluid model theorists.\cite{2-Peres-1995-FM,1-Sigmond-1997-EXP,2-Morrow-1997-FM,2-Donko-1999-FM,2-Akishev-1999-FM,2-Allen-2007-FM} 

Does the aforementioned positive feedback mechanism, which is asserted
in the fluid models, represent the fundamental mechanism behind the
phenomenon of self-sustained oscillations in the DC glow discharges?
This is a subtle question because I have shown recently that an ion
confined between the DC voltage biased plane-parallel conductors goes
through an undamped self-oscillatory motion.\cite{4-Cho-1} Such self-oscillatory
motion requires that an ion is electrically polarizable, but it does
not necessarily involve or require the discussed positive feedback
mechanism which is asserted in the fluid models. The fact that an
ion must be electrically polarized excludes electrons from consideration
in the discussion of self-oscillations presented in this work. However,
the proton, which is known to be electrically polarizable,\cite{proton-polarizability}
is expected to self-oscillate in the DC glow discharges according
to this alternative theory. 

Remarkably, the predictions of this alternative theory\cite{4-Cho-2}
qualitatively agree with the predictions made by the fluid models.\cite{2-Morrow-1997-FM,2-Akishev-1999-FM}
Both theories predict a saw-tooth shaped waveform for the induced
currents in the electrodes. Furthermore, sharp pulses of radiation
output are predicted by both theories to accompany the abrupt rises
in the induced electrode currents. Such remarkable similarities in
the predictions by the two very dissimilar theories suggest that the
discussed positive feedback mechanism, which is asserted in the fluid
models, may not necessarily represent the fundamental mechanism behind
the phenomenon of self-sustained oscillations in the DC glow discharges. 

In this work, I shall present some new aspects of this alternative
theory\cite{4-Cho-2} for further exploitation of the phenomenon of
self-sustained ion oscillations in the DC glow discharges. These theoretical
predictions are explicitly compared with various experimental results
for the validation of the model. By direct application of the model
to describe the charged-particle oscillations in dusty (or complex)
plasmas experiments, I shall reveal that the phenomenon of self-sustained
oscillations in both dusty plasmas and DC glow discharges involves
a common physical mechanism in which self-oscillations are attributed
to the interaction between an ionized particle and the surface charges
induced by it at the surfaces of the particle confining electrodes.
Considering that particle oscillations in both dusty plasmas and DC
glow discharges experiments involve ionized particles that are only
differ in their physical sizes and masses, such outcome is not too
surprising. After all, apart from this, everything else is nearly
identical from the physics point of view in both experiments.

\section{Brief outline of the theory}

The self-sustained oscillations in DC glow discharges are now discussed
briefly in the framework of presented alternative theory using the
configuration illustrated in Fig. \ref{fig:1}. The electric potential
between the plates is obtained by solving the Laplace equation with
appropriate boundary conditions.\cite{4-Cho-1} With this electric
potential, induced surface charges at the surfaces of conductor plates
are obtained by application of the Gauss's law. Such charge distributions
act on an ion and generate resultant force given by $\mathbf{F}_{T}=\mathbf{F}_{1}+\mathbf{F}_{2}-\mathbf{e}_{z}mg,$
where $m$ is the mass of an ion, $g=9.8\,\textnormal{m}/\textnormal{s}^{2}$
is the gravitational constant, and $\mathbf{F}_{1}$ ($\mathbf{F}_{2}$)
is the force between the ion and the surface charges induced by it
at the surface of anode (cathode). The explicit forms of $\mathbf{F}_{1}$
and $\mathbf{F}_{2}$ are given by\cite{4-Cho-1} 
\[
\mathbf{F}_{1}=\mathbf{e}_{z}\pi\epsilon_{0}\kappa_{3}\nu\left\{ \frac{\nu}{4s^{2}}+E_{p}\left[\frac{\gamma\left(b^{3}-a^{3}\right)-b^{3}}{4s^{3}}-1\right]\right\} ,
\]
 
\[
\mathbf{F}_{2}=-\mathbf{e}_{z}\pi\epsilon_{0}\kappa_{3}\nu\left\{ \frac{\nu}{4\left(h-s\right)^{2}}-E_{p}\left[\frac{\gamma\left(b^{3}-a^{3}\right)-b^{3}}{4\left(h-s\right)^{3}}-1\right]\right\} ,
\]
 where $\epsilon_{0}$ is the vacuum permittivity, $E_{p}\equiv\left\Vert \mathbf{E}_{p}\right\Vert $
is the parallel-plate electric field strength, the parameter $s$
is the distance from the ion's physical center to the anode; and,
the terms $\gamma$ and $\nu$ are defined as 
\begin{equation}
\gamma=\frac{3\kappa_{3}b^{3}}{\left(\kappa_{2}+2\kappa_{3}\right)b^{3}+2\left(\kappa_{2}-\kappa_{3}\right)a^{3}},\label{eq:gama}
\end{equation}
 
\begin{equation}
\nu=\frac{2a\left(b-a\right)\sigma_{1}}{\epsilon_{0}\kappa_{2}}+\frac{a^{2}\sigma_{1}+b^{2}\sigma_{2}}{\epsilon_{0}\kappa_{3}},\label{eq:nu}
\end{equation}
 where $\sigma_{1}$ ($\sigma_{2}$) is the surface charge density
at the ion's core (shell) of radius $r=a$ ($r=b$), and the dielectric
constants $\kappa_{2}$ and $\kappa_{3}$ are depicted in Fig. \ref{fig:1}.
The resultant force on an ion is, hence, given by \begin{widetext}
\begin{equation}
\mathbf{F}_{T}\left(z_{d}\right)=\mathbf{e}_{z}\frac{\pi\epsilon_{0}\kappa_{3}\nu}{4}\left\{ \frac{\nu}{\left(z_{d}+b\right)^{2}}-\frac{\nu}{\left(h-z_{d}-b\right)^{2}}+E_{p}\left[\frac{\gamma\left(b^{3}-a^{3}\right)-b^{3}}{\left(z_{d}+b\right)^{3}}+\frac{\gamma\left(b^{3}-a^{3}\right)-b^{3}}{\left(h-z_{d}-b\right)^{3}}-8\right]\right\} -\mathbf{e}_{z}mg,\label{eq:FT(zd)}
\end{equation}
 where $s=z_{d}+b$ in $\mathbf{F}_{1}$ and $\mathbf{F}_{2}$ (see
Fig. \ref{fig:1}). The potential energy associated with this force
is given by\cite{4-Cho-2} 

\begin{align}
U\left(z_{d}\right) & =\frac{\pi\epsilon_{0}\kappa_{3}\nu}{4}\left\{ \frac{\nu}{z_{d}+b}+\frac{\nu}{h-z_{d}-b}-\frac{4\nu}{h}+E_{p}\left[\frac{\gamma\left(b^{3}-a^{3}\right)-b^{3}}{2\left(z_{d}+b\right)^{2}}-\frac{\gamma\left(b^{3}-a^{3}\right)-b^{3}}{2\left(h-z_{d}-b\right)^{2}}+8\left(z_{d}+b-\frac{h}{2}\right)\right]\right\} \nonumber \\
 & +mg\left(z_{d}+b-\frac{h}{2}\right);\label{eq:U(zd)}
\end{align}
 and, the equation of motion associated with this force, i.e., Eq.
(\ref{eq:FT(zd)}), can be expressed as 
\begin{equation}
\frac{d^{2}z_{d}}{dt^{2}}=\frac{\pi\epsilon_{0}\kappa_{3}\nu}{4m}\left\{ \frac{\nu}{\left(z_{d}+b\right)^{2}}-\frac{\nu}{\left(h-z_{d}-b\right)^{2}}+E_{p}\left[\frac{\gamma\left(b^{3}-a^{3}\right)-b^{3}}{\left(z_{d}+b\right)^{3}}+\frac{\gamma\left(b^{3}-a^{3}\right)-b^{3}}{\left(h-z_{d}-b\right)^{3}}-8\right]\right\} -g,\label{eq:ODE-newton}
\end{equation}
 \end{widetext} where $\gamma$ and $\nu$ are defined in Eqs. (\ref{eq:gama})
and (\ref{eq:nu}), respectively.\cite{4-Cho-1,4-Cho-2}

\section{Comparison to the experiments}

\begin{figure}[h]
\begin{centering}
\includegraphics[width=0.9\columnwidth]{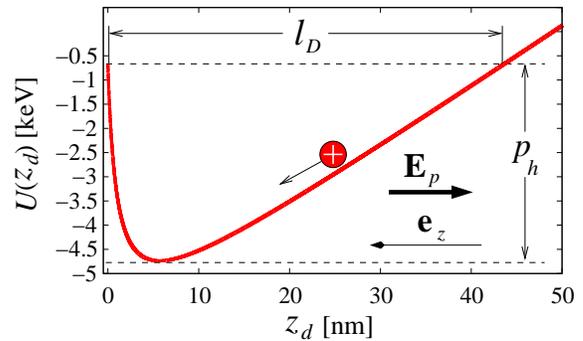}
\par\end{centering}

\caption{Potential energy is plotted for $V_{T}=100\,\textnormal{V}$ using
parameter values defined in Eq. (\ref{eq:parameter-values}). Here,
$V_{T}=100\,\textnormal{V}$ corresponds to $E_{p}=1\,\textnormal{GV}/\textnormal{m}.$
\label{fig:2} }
\end{figure}

Gases under pressure in general, including noble gases, are composed
of atomic clusters, a state of matter between molecules and solids,
due to Van der Waals interaction.\cite{argon-cluster-1956,argon-cluster-1977,argon-cluster-2009}
In gaseous argon, each spherical atomic clusters of radius $r\approx1\,\textnormal{nm}$
contains roughly $135$ argon atoms.\cite{argon-cluster-1977} Modeling
particles in the gas as atomic clusters is important because gases
are delivered to the laboratories in pressurized containers, in which
environment individual particles in gas exist in the form of atomic
clusters. That said, I shall explicitly work with an argon atomic
cluster of radius $r\approx1\,\textnormal{nm},$ which for brevity
is simply referred to as ``ion'' hereafter. Such ion is not expected
to be core-shell structured like the one depicted in Fig. \ref{fig:1}.
The shell portion of an ion can be eliminated by choosing $b=a,$
$\kappa_{2}=\infty,$ and $\sigma_{2}=0\,\textnormal{C}/\textnormal{m}^{2}.$
For a positive ion, its surface charge density is given by $\sigma_{1}=Nq_{e}/\left(4\pi a^{2}\right),$
where $N$ is the number of electrons removed from the particle and
$q_{e}=1.602\times10^{-19}\,\textnormal{C}$ is the fundamental charge
magnitude. Without loss of generality, and for the purpose of clear
illustration in this work, I shall choose $N=250$ and $a=1\,\textnormal{nm}.$
This corresponds to the ion's surface charge density of $\sigma_{1}\approx3.19\,\textnormal{C}/\textnormal{m}^{2}.$
Assuming the mass of an argon atom is $6.63\times10^{-26}\,\textnormal{kg},$
a spherical atomic cluster composed of $135$ argon atoms has a total
mass of $m\approx8.95\times10^{-24}\,\textnormal{kg},$ where the
masses of missing $N$ electrons have been neglected. Purely for convenience,
I shall assume that the cathode is grounded and the space between
the two conductor plates in Fig. \ref{fig:1} is a vacuum with a gap
of $h=100\,\textnormal{nm}.$ The obtained parameter values are summarized
here for reference: 
\begin{equation}
\left\{ \begin{array}{c}
\kappa_{2}=\infty,\;\;\kappa_{3}=1,\;\; h=100\,\textnormal{nm},\;\; V_{L}=0\,\textnormal{V},\\
\sigma_{1}\approx3.19\,\textnormal{C}/\textnormal{m}^{2},\;\;\sigma_{2}=0\,\textnormal{C}/\textnormal{m}^{2},\\
m\approx8.95\times10^{-24}\,\textnormal{kg},\;\; b=a=1\,\textnormal{nm}.
\end{array}\right.\label{eq:parameter-values}
\end{equation}
 Illustrated in Fig. \ref{fig:2} is a plot of the potential energy
well, where Eq. (\ref{eq:U(zd)}) has been plotted for $V_{T}=100\,\textnormal{V}$
using the parameters defined in Eq. (\ref{eq:parameter-values}).
The width $l_{D}$ of the potential energy well decreases with an
increase in the electric field strength $E_{p}.$ Such characteristic
provides an elegant explanation to the experimental observations in
which the oscillation frequencies increase with the applied electric
field strength. Fox seems to be the first to discuss on such characteristic
using a parabolic potential model.\cite{0-Fox-1931} 

In the case of positive ions, the potential energy well gets formed
in vicinity of the anode at the presence of applied static electric
field. On the other hand, for the negative ions, the same electric
field results in formation of the potential energy well in vicinity
of the cathode. Such characteristic provides an elegant explanation
to the phenomenon of fireball formation in the laboratory plasma.
When a positively biased electrode is immersed in plasma, a region
of intense glow appears in vicinity of the electrode. Such glow formation
is referred to as a plasma fireball.\cite{fireball} In the framework
of this alternative theory\cite{4-Cho-1,4-Cho-2} elaborated in this
work, the formation of such plasma fireballs can be elegantly explained
from the  potential energy well which gets formed near the anode for
positive ions. Since the dynamics of a plasma involves collective
motions of its constituent atoms, a weakly ionized plasma fireball,
as a whole, self-oscillates near the electrode as if it were a weakly
ionized single superparticle. Throughout this work, I shall use the
term ``superparticle'' to refer to such entity as plasma fireball
whose dynamics can be represented by an equivalent single-particle
picture. In the equivalent single-particle picture, the self-oscillation
dynamics of a plasma fireball is described by the equation of motion
defined in Eq. (\ref{eq:ODE-newton}) with appropriate effective mass
and surface charge densities prescribed. With that in mind, the plasma
fireball effectively has a very large mass and carries a net charge
which is very weak. Consequently, induced current oscillations of
relatively low frequencies and small amplitudes are expected to be
generated at the anode near a self-oscillating plasma fireball. Such
prediction is consistent with the experimental observation by Stenzel
\textit{et al}.\cite{fireball} Notice that this alternative theory
can be directly applied to describe the self-oscillations of ionized
dust particles in dusty plasmas. It is remarkable that the profile
of the potential energy well illustrated in Fig. \ref{fig:2} qualitatively
agrees with the measurements by Tomme \textit{et al}.\cite{dusty-plasma-Tomme-sheat-potential}
and Arnas \textit{et al}.\cite{dusty-plasma-Arnas - 2000} in dusty
plasmas. Further discussion on this is provided in the sections A
thru C of the appendix.

\begin{figure*}[t]
\begin{centering}
\includegraphics[width=0.9\columnwidth]{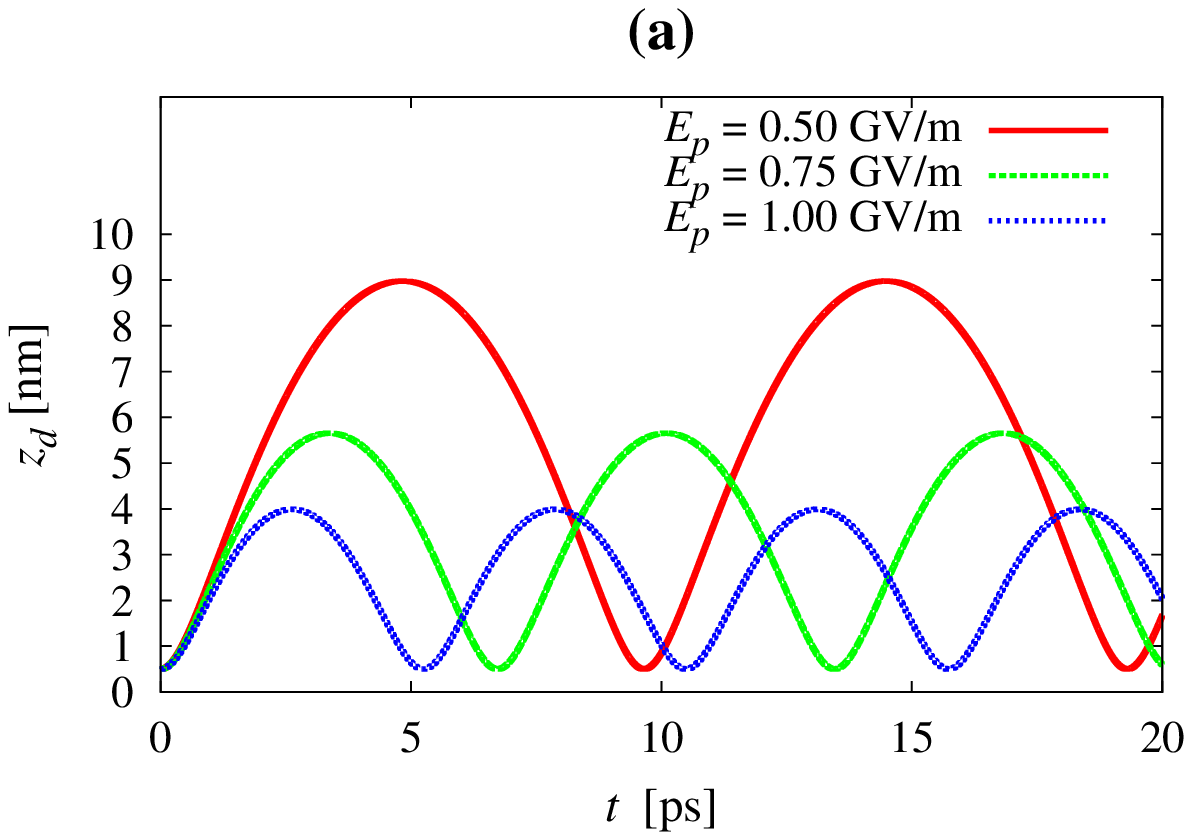}\includegraphics[width=0.9\columnwidth]{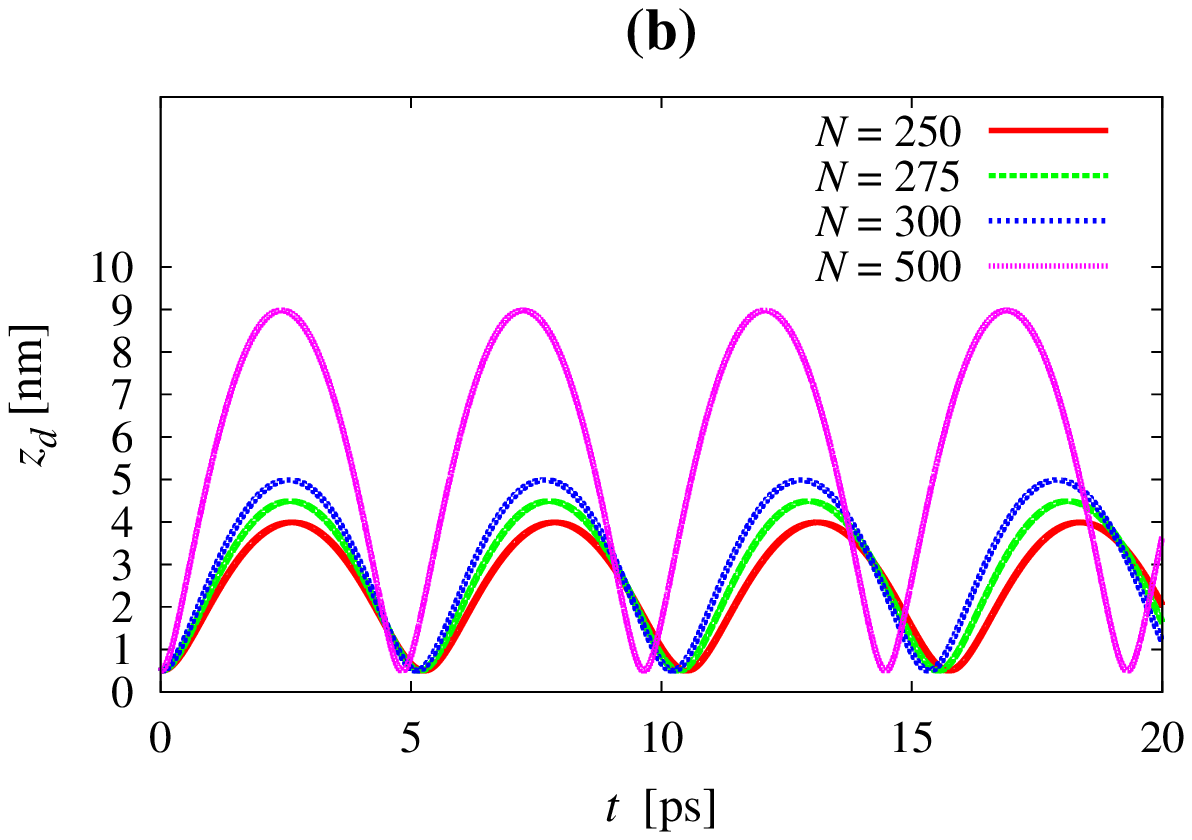}
\par\end{centering}

\begin{centering}
\includegraphics[width=0.9\columnwidth]{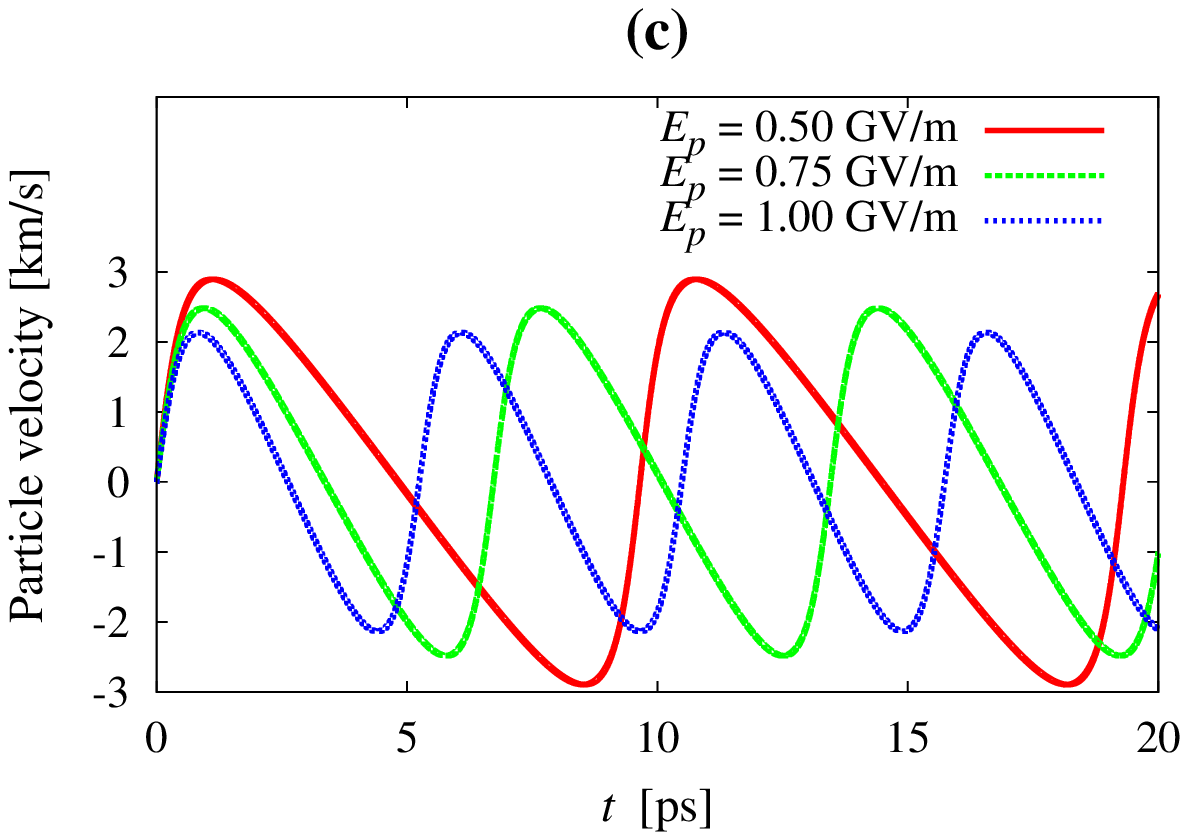}\includegraphics[width=0.9\columnwidth]{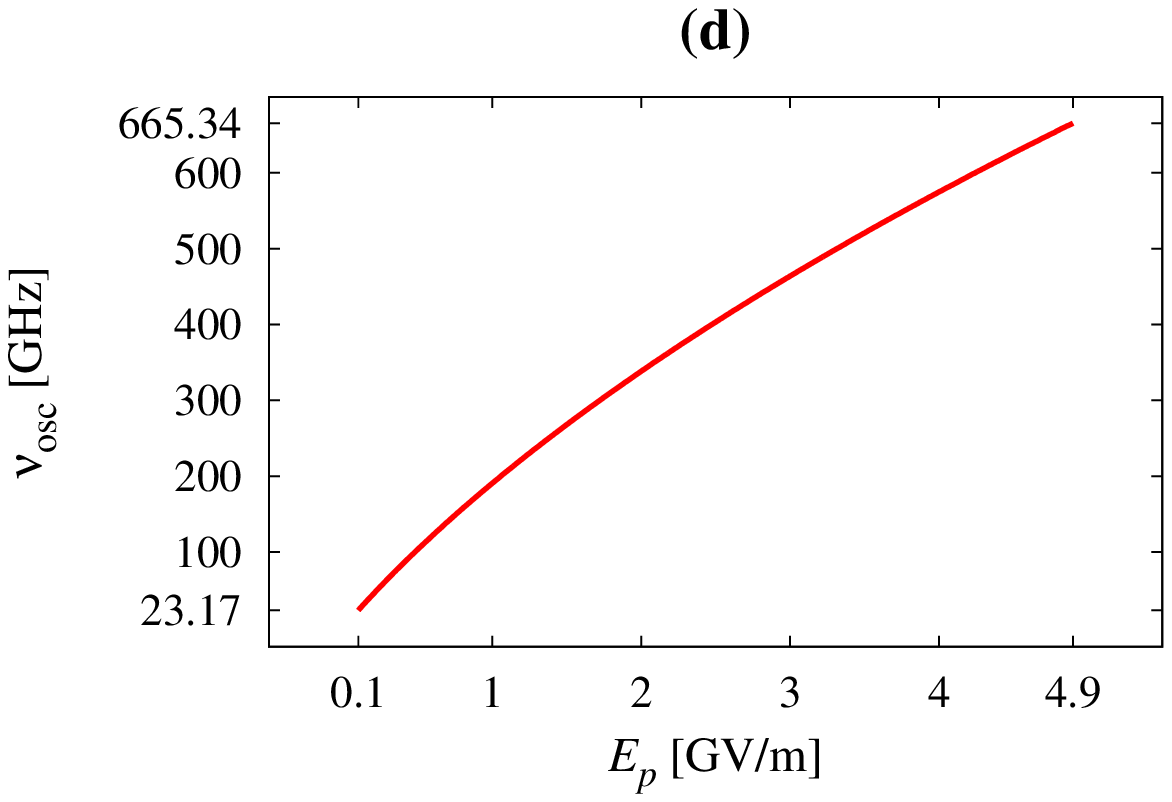}
\par\end{centering}

\caption{(Color online) Equation (\ref{eq:ODE-newton}) is plotted, where in
(a) electric field is varied and in (b) ion's surface charge density
$\sigma_{1}=Nq_{e}/\left(4\pi a^{2}\right)$ is varied at $E_{p}=1\,\textnormal{GV}/\textnormal{m}.$
(c) Ion velocity corresponding to the plot in (a). (d) Dependence
of oscillation frequency $\nu_{\textnormal{osc}}$ on electric field
strength $E_{p}.$ Similar plot for the lower frequencies involving
weakly charged ions subject to smaller $E_{p}$ is also provided in
the section E of appendix.  In (a,b,c,d), all of the unspecified parameter
values are from Eq. (\ref{eq:parameter-values}) and the initial conditions
are from Eq. (\ref{eq:IC-positive}). \label{fig:3}}
\end{figure*}

The particle's equation of motion described in Eq. (\ref{eq:ODE-newton})
can be solved via Runge-Kutta method using the parameter values defined
in Eq. (\ref{eq:parameter-values}) and the initial conditions given
by 
\begin{equation}
z_{d}\left(0\right)=0.5\,\textnormal{nm}\quad\textnormal{and}\quad\frac{dz_{d}\left(0\right)}{dt}=0\,\frac{\textnormal{nm}}{\textnormal{s}}.\label{eq:IC-positive}
\end{equation}
 The results for $V_{T}=50\,\textnormal{V},$ $75\,\textnormal{V},$
and $100\,\textnormal{V}$ are shown in Fig. \ref{fig:3}(a), where
it shows the oscillation frequencies increasing with the electric
field strength. The oscillation frequencies also increase with the
ion's surface charge density, but at the rate which is significantly
slower than it does with the electric field strength (see Fig. \ref{fig:3}(b)).
For instance, when the electric field strength is doubled from $E_{p}=0.5\,\textnormal{GV}/\textnormal{m}$
to $E_{p}=1\,\textnormal{GV}/\textnormal{m},$ oscillation frequency
nearly doubles from $\nu_{\textnormal{osc}}\approx104\,\textnormal{GHz}$
to $\nu_{\textnormal{osc}}\approx191\,\textnormal{GHz}.$ However,
doubling the ion's surface charge density, i.e., $\sigma_{1}=Nq_{e}/\left(4\pi a^{2}\right),$
from $N=250$ to $N=500$ only slightly increases the oscillation
frequency from $\nu_{\textnormal{osc}}\approx191\,\textnormal{GHz}$
to $\nu_{\textnormal{osc}}\approx207\,\textnormal{GHz}.$ Such characteristics
is consistent with the observations by Fox\cite{0-Fox-1931} and Bošan
\textsl{et} \textsl{al}.,\cite{Bosan-1988} where they have reported
that oscillation frequencies do not seem to be very dependent on the
type of ions used in the discharge. It is well known that dissimilar
atomic gases have different ionization tendencies. Based on this,
it can be inferred that $\sigma_{1},$ in general, for different atoms
under identical conditions are different. 

Although a relatively high static electric field is being applied
to the ion  in Fig. \ref{fig:3}(a), its kinetic energies are small
due to the fact that ion is going through frequent turnings inside
a tight potential energy well. The ion's kinetic energy decreases
with increased electric field strength for that matter and this is
illustrated in Fig. \ref{fig:3}(c). At the maximum speed of $\sim2.912\,\textnormal{km}/\textnormal{s},$
the ion  gains kinetic energy of $\mathcal{E}_{\textnormal{K}}\approx237\,\textnormal{eV},$
which is small compared to the depth of $p_{h}\approx4.1\,\textnormal{keV}$
for the potential energy well but large compared to the average room
temperature thermal energy of $\mathcal{E}_{\textnormal{T}}=1.5k_{\textnormal{B}}T\approx39\,\textnormal{meV}$
(here, $T=300\,\textnormal{K}$ and $k_{\textnormal{B}}=8.62\times10^{-5}\,\textnormal{eV}/\textnormal{K}$
is the Boltzmann constant). Self-sustained oscillations in the DC
glow discharges can thus be expected to have high thermal stability. 

In the experiments, self-sustained oscillations of ions in the DC
glow discharges occur only in the negative resistance region of the
voltage-current characteristic curve.\cite{1-Petrovic-1997-EXP,1-Lozneanu-2002-EXP,3-Phelps-1992-ECM,2-Donko-1999-FM}
Such negative resistance region in the voltage-current characteristic
curve is characterized by $V_{\textnormal{th1}}\leq V_{T}\leq V_{\textnormal{th2}},$
where $V_{\textnormal{th1}}$ and $V_{\textnormal{th2}}$ are some
threshold voltages. Equation (\ref{eq:ODE-newton}) provides an elegant
explanation to such properties. For instance, Eq. (\ref{eq:ODE-newton})
yields solutions that are unphysical for $E_{p}<E_{p,\textnormal{th1}}$
and $E_{p}>E_{p,\textnormal{th2}},$ where $E_{p,\textnormal{th1}}$
and $E_{p,\textnormal{th2}}$ are the threshold electric field strengths.
The physical solutions are only obtained for $E_{p,\textnormal{th1}}\leq E_{p}\leq E_{p,\textnormal{th2}}.$
It is difficult to pinpoint $E_{p,\textnormal{th1}}$ and $E_{p,\textnormal{th2}}$
without the analytical solution of Eq. (\ref{eq:ODE-newton}) at hand.
Nevertheless, these can be roughly estimated on the grounds of physicality.
To illustrate this, the oscillation frequency is first plotted as
a function of the electric field strength in Fig. \ref{fig:3}(d)
by numerically solving Eq. (\ref{eq:ODE-newton}) using the parameter
values and the initial conditions from Eqs. (\ref{eq:parameter-values})
and (\ref{eq:IC-positive}), respectively. The solutions obtained
for $E_{p}\lesssim0.1\,\textnormal{GV}/\textnormal{m}$ and $E_{p}\gtrsim4.9\,\textnormal{GV}/\textnormal{m}$
are unphysical. For instance, in the case of $E_{p}\gtrsim4.9\,\textnormal{GV}/\textnormal{m},$
solutions show that the ion penetrates into the anode's surface whereas
for $E_{p}\lesssim0.1\,\textnormal{GV}/\textnormal{m},$ solutions
yield peak to peak amplitudes that are larger than $h/2.$ This latter
case, although mathematically allowed, is unphysical because, for
a positive ion, oscillations can only exist for $z_{d}<h/2-b$ from
the argument based on the grounds of physicality.\cite{4-Cho-1} Hence,
the physical oscillatory solutions exist only for $0.1\,\textnormal{GV}/\textnormal{m}\lesssim E_{p}\lesssim4.9\,\textnormal{GV}/\textnormal{m},$
where the upper and the lower bounds of $E_{p}$ are only rough estimates
based on the grounds of physicality arguments. Further discussion
on the properties of the physical and the unphysical oscillatory solutions
are provided in the section D of the appendix. 

Recently, Lotze \textsl{et} \textsl{al}.\cite{Science-Lotze} reported
on an indirect account of oscillations involving a single $\textnormal{H}_{\textnormal{2}}$
molecule in the DC voltage biased conductors near an absolute zero
temperature of $T=5\,\textnormal{K}.$ According to their findings,
the $\textnormal{H}_{\textnormal{2}}$ molecule in the junction, which
is the space between the atomically clean $\textnormal{Cu}\left(111\right)$
surface and the STM (scanning tunneling microscope) tip mounted on
a cantilever, self-oscillates between the two unknown positional states
when a threshold DC bias voltage is applied to the electrodes, i.e.,
the copper $\textnormal{Cu}\left(111\right)$ surface and the STM
tip. The footprint of self-sustained oscillations is the presence
of the negative differential conductance  in their measurement. Gupta
\textsl{et} \textsl{al}.\cite{Science-Phys-Rev-Gupta} showed that
such negative differential conductance is due to the $\textnormal{H}_{\textnormal{2}}$
molecules in the junction. In the DC glow discharges, self-sustained
oscillations arise as a consequence of the negative differential resistance.\cite{1-Petrovic-1997-neg-resistance,1-Petrovic-1997-EXP,1-Lozneanu-2002-EXP}
Since the negative differential conductance and the negative differential
resistance are reciprocally related, they represent an equivalent
description of the same physical processes. It is well known that
the $\textnormal{H}_{\textnormal{2}}$ molecules can be ionized by
a process of an electron impact.\cite{electron-impact-2010} When
a threshold DC bias voltage is applied to the electrodes, the energetic
electrons begin to tunnel through the junction. It is possible that
the $\textnormal{H}_{\textnormal{2}}$ molecules in the junction are
ionized in the process. If that is indeed the case, what Lotze \textsl{et}
\textsl{al}. reported may be an indirect account of self-sustained
oscillations involving a single or a few ions in the DC glow corona.\cite{4-Cho-2}
Their observation can serve as a catalyst for the future experiments
in the few-particle DC plasma experiments. 

\begin{figure}[h]
\begin{centering}
\includegraphics[width=0.9\columnwidth]{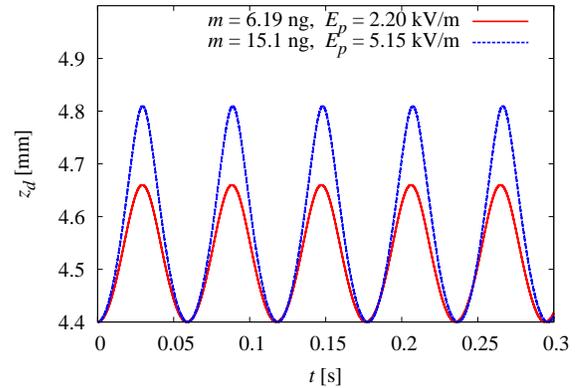}
\par\end{centering}

\caption{(Color online) Equation (\ref{eq:ODE-newton}) is plotted using parameter
values and initial conditions defined in Eqs. (\ref{eq:Arnas-parameter}).
Oscillation frequency of $\nu_{\textnormal{osc}}\approx17\,\textnormal{Hz}$
is predicted by the theory, which is consistent with the measurement
by Arnas \textit{et al}.\cite{dusty-plasma-Arnas - 1999} In the plot,
the anode is located at $z_{d}=0\,\textnormal{mm}$ and the cathode
is located at $z_{d}=5\,\textnormal{mm}.$\label{fig:4}}
\end{figure}

Arnas \textit{et al}.\cite{dusty-plasma-Arnas - 1999} reported that
a dust particle made of hollow glass microsphere of radius $r\approx32\,\mu\textnormal{m}$
with mass density of $\rho_{m}\approx110\,\textnormal{kg}/\textnormal{m}^{3}$
and carrying a total electrical charge of $Q\approx-4.3\times10^{5}q_{e},$
where $q_{e}=1.602\times10^{-19}\,\textnormal{C},$ self-oscillates
at a frequency of $\nu_{\textnormal{osc}}\approx17\,\textnormal{Hz}$
inside the plasma sheath. The plasma sheath environment, in many respects,
is similar to the empty space region between the DC voltage biased
plane-parallel conductors. Ionized particle oscillations in the plasma
sheath can therefore be modeled from the simple apparatus illustrated
in Fig. \ref{fig:1}. That said, the parameters and the initial conditions
in the experiment by Arnas \textit{et al}. can be summarized as follow:
\begin{equation}
\left\{ \begin{array}{c}
\kappa_{2}=6.5,\;\;\kappa_{3}=1,\;\; h=5\,\textnormal{mm},\;\; E_{p}=5.15\,\textnormal{kV}/\textnormal{m},\\
\sigma_{1}=0\,\textnormal{C}/\textnormal{m}^{2},\;\;\sigma_{2}=-5.35\,\mu\textnormal{C}/\textnormal{m}^{2},\\
m\approx15.1\,\textnormal{ng},\;\; a=0\,\textnormal{m},\;\; b=32\,\mu\textnormal{m},\\
z_{d}\left(0\right)=4.4\,\textnormal{mm},\;\;\dot{z}_{d}\left(0\right)=0\,\textnormal{mm}/\textnormal{s},\;\;\dot{z}_{d}\equiv dz_{d}/dt,
\end{array}\right.\label{eq:Arnas-parameter}
\end{equation}
 where the details of how these were obtained are explained in the
section B of the appendix. Using these, the equation of motion described
in Eq. (\ref{eq:ODE-newton}) can be solved numerically via Runge-Kutta
method. The result is shown in Fig. \ref{fig:4}. One can readily
verify that the particle oscillates at a frequency of approximately
$\nu_{\textnormal{osc}}\approx17\,\textnormal{Hz},$ which is consistent
with the measurement by Arnas \textit{et al}.\cite{dusty-plasma-Arnas - 1999}

Although Arnas \textit{et al}. explicitly states that their glass
microsphere is hollow structured, their paper does not provide any
information regarding the radius of the void inside of their hollow
glass microsphere. Due to the lack of this information, the mass of
an hollow glass microsphere in Eq. (\ref{eq:Arnas-parameter}) has
been computed assuming a solid microsphere. Such assumption overestimates
the mass of an hollow glass microsphere. In fact, in their paper,
Arnas \textit{et al}. indicates the gravitational force of $F_{g}=\left(1.5\pm0.3\right)\times10^{-10}\,\textnormal{N}$
for their hollow glass microsphere.\cite{dusty-plasma-Arnas - 1999}
Such force corresponds to a mass of $m\approx15.3\,\textnormal{ng},$
which is exactly the mass computed assuming a solid glass microsphere
in Eq. (\ref{eq:Arnas-parameter}). If instead a smaller mass of $m\approx6.19\,\textnormal{ng}$
is assumed, an electric field strength of $E_{p}=2.2\,\textnormal{kV}/\textnormal{m}$
is sufficient to generate an oscillation frequency of $\nu_{\textnormal{osc}}\approx17\,\textnormal{Hz}.$
This value for the electric field strength, i.e., $E_{p}=2.2\,\textnormal{kV}/\textnormal{m},$
is consistent with the local electric field strength in the plasma
sheath obtained by Arnas \textit{et al}. in their dusty plasma experiment.
The particle's oscillatory motion corresponding to this latter case,
where $m\approx15.1\,\textnormal{ng}$ and $E_{p}=5.15\,\textnormal{kV}/\textnormal{m},$
has been plotted in Fig. \ref{fig:4} for comparison.

This remarkable result demonstrates that self-sustained oscillations
in both dusty plasmas and DC glow discharges share a common physical
mechanism, in which the oscillations can be attributed to the interaction
between the ionized particle and the surface charges induced by it
at the bounding electrodes. Such outcome makes sense because the oscillating
ionized particles in both dusty plasmas and DC glow discharges experiments
are only differ in their physical sizes and masses. Apart from this,
the two systems are nearly identical in nature from the physics point
of view. 

Quite often in dusty plasmas, theoretical models based on simple harmonic
excitations of small amplitude oscillations are employed for the description
of the particle dynamics. One such model is given by\cite{dusty-plasma-Fortov-2004}
\begin{equation}
\frac{d^{2}z}{dt^{2}}+\nu_{\textnormal{dn}}\frac{dz}{dt}+\Omega_{\textnormal{v}}^{2}z=\frac{f_{0}\cos\omega t}{m_{\textnormal{d}}},\label{eq:SHM-dusty-plasma}
\end{equation}
 where $z$ is the particle displacement, $m_{\textnormal{d}}$ is
the particle's mass, $\omega$ is the angular frequency of oscillation,
$t$ is the time, and $f_{0}$ is the amplitude of the external force.
Particle's equation of motion based on such model, Eq. (\ref{eq:SHM-dusty-plasma}),
is limited to the descriptions of oscillations involving a small amplitude
displacements about the stationary equilibrium position. Besides this
limitation, the equation of motion described by Eq. (\ref{eq:SHM-dusty-plasma}),
as it stands, cannot be directly applied to describe the particle's
oscillatory motion due to the fact that the quantities $\nu_{\textnormal{dn}}$
and $\Omega_{\textnormal{v}}$ must be first determined experimentally.\cite{dusty-plasma-Melzer-1994}
Consequently, Eq. (\ref{eq:SHM-dusty-plasma}) is only useful when
determining the total charge carried by the particle. Although Eq.
(\ref{eq:SHM-dusty-plasma}) provides an indirect method of measuring
the particle's total charge, as it stands, it says nothing about the
fundamental mechanism behind the self-sustained oscillations of ions
in dusty plasmas. 

The equation of motion described in Eq. (\ref{eq:ODE-newton}) is
distinguished from the one illustrated in Eq. (\ref{eq:SHM-dusty-plasma})
in that its physical description is not limited to just small amplitude
oscillations but covers particle oscillations of any amplitude ranges.
Further, Eq. (\ref{eq:ODE-newton}) does not depend on quantities
like $\nu_{\textnormal{dn}}$ and $\Omega_{\textnormal{v}},$ which
terms must be determined experimentally. Instead, Eq. (\ref{eq:ODE-newton})
completely describes the dynamics of charged-particle motion solely
based on the information obtained from the particle's physical properties
(i.e., mass, surface charge density, and dielectric constant, etc.)
and the local electric field strength. In the case where the mass
of an ionized particle is represented by that of an ionized atom,
Eq. (\ref{eq:ODE-newton}) describes the dynamics of self-oscillating
ionized atom in the DC glow discharges. In the previous work,\cite{4-Cho-2}
I have discussed that Eq. (\ref{eq:ODE-newton}) yields results that
qualitatively agree with certain aspects of self-oscillations in the
DC glow corona experiments predicted by theories based on the fluid
models.\cite{2-Morrow-1997-FM,2-Akishev-1999-FM} If plasma in the
DC glow corona experiment is effectively treated as a self-oscillating
weakly ionized single superparticle, Eq. (\ref{eq:ODE-newton}) provides
a satisfactory description of the electrode current oscillations in
the DC glow corona. Further discussion on this is provided in the
section E of the appendix. 

Until now, no single theory was able to successfully explain self-sustained
oscillations in both dusty plasmas and DC glow discharges experiments.
For years, experimental evidences hinted that these were related phenomena,
but without any definitive conclusions. In this respect, Eq. (\ref{eq:ODE-newton})
is the first theoretical model to provide such definitive conclusions.

\section{Device application}

It is well known that oscillating ions generate electric dipole radiation.
Such property can be utilized to develop a wideband electromagnetic
radiation source in which the frequency of emitted radiation can be
tuned by varying the DC bias voltage across the electrodes.\cite{4-Cho-1,4-Cho-2}
According to Zouche and Lefort, the plane-parallel plate electrodes
made of nickel-silver composite material, which are separated by a
gap of $h=100\,\textnormal{nm},$ can support DC bias voltages up
to $\sim400\,\textnormal{V}$ across the electrodes before an onset
of electrical breakdown.\cite{7-IEEE-vacuum} With improvements in
the electrical breakdown characteristics, such device can be engineered
to cover both the microwave and the terahertz band of electromagnetic
spectrum with high efficiency. 

The first experimental evidence that the phenomenon of self-sustained
ion oscillations in the DC glow discharges has a potential applications
in the development of a wideband electromagnetic radiation source
came from McClure in 1963, when he reported an oscillation frequency
of $\nu_{\textnormal{osc}}\approx20\,\textnormal{MHz}$ in a low pressure
glow discharge tube, which comprised of cylindrical hollow cathode
and a very thin coaxial wire anode.\cite{McClure} Unfortunately,
McClure provided no theoretical model to explain his finding. Such
concentric cylinder configuration for the electrodes is a typical
apparatus found in many DC glow discharges experiments. Nearly two
decades later, in 1980, Alexeff and Dyer shrunk the aforementioned
concentric cylinder configuration (filled with air at gas pressure
of $0.1\,\textnormal{mTorr}$) to the size of a pen and successfully
demonstrated the generation of microwave radiations in the gigahertz
frequency ranges.\cite{Orbitron-Alexeff-1980} Their pen sized coaxial
configuration was later reported to generate radiation at terahertz
frequency of $\nu_{\textnormal{osc}}\approx1\,\textnormal{THz}$ with
an output radiation power of $P_{rad}\approx1.5\,\textnormal{W}.$\cite{Orbitron-Alexeff-1985,Orbitron-Alexeff-1989}
In order to explain their observations, Alexeff and Dyer proposed
a model which they referred to as the \textit{orbitron theory}.\cite{Orbitron-Alexeff-1980,Orbitron-Alexeff-1985,Orbitron-Alexeff-1989}
The basic idea behind the orbitron theory is very simple. The electrons
emitted from the inner surface of the outer concentric cylinder (which
part represents the cathode in the coaxial configuration) orbit around
a thin coaxial wire anode; and, such electron orbital motion generates
the detected high frequency electromagnetic radiation. However, various
experimental results reported by other groups\cite{Orbitron-BAD-Schumacher-Harvey-1984,Orbitron-BAD-Felsteiner-1987,Orbitron-BAD-Stenzel-1988}
contradicted the orbitron model; and, the orbitron theory is no longer
considered as the correct description of the physics behind the observations
reported by Alexeff and Dyer. 

Somewhat similar to the aforementioned pen sized coaxial configuration
investigated by Alexeff and Dyer is the microhollow cathode discharges
(MHCDs) configuration, which was first introduced by Schoenbach \textit{et
al}.\cite{1-Schoenbach-1997-EXP} However, unlike the coaxial\textit{
} configuration, where the self-pulsing frequencies of ions typically
lie in the gigahertz ranges, the self-pulsing of ions in the DC glow
discharges for typical MHCDs configurations lie only in some tens
of kilohertz frequency ranges.\cite{3-He-2012-ECM,MHC-Rousseau-2006,MHC-Aubert-2007}
Why? The answer to such discrepancy lies in the differences in the
gas pressures applied in two configurations. In the high frequency
design by Alexeff and Dyer, the pen sized coaxial device is filled
with gas at a very low gas pressure of $0.1\,\textnormal{mTorr}$
whereas, in typical MHCDs configurations, the device is maintained
at gas pressures on the order of tens (or hundreds) of torrs. For
instance, in the MHCDs experiment by Aubert \textit{et al}.,\cite{MHC-Rousseau-2006,MHC-Aubert-2007}
the gas pressure ranged from $40$ to $200\,\textnormal{Torr.}$ He
\textit{et al}.\cite{3-He-2012-ECM} worked with somewhat lower gas
pressure of $\sim1\,\textnormal{Torr}$ (i.e., $133\,\textnormal{Pa}$)
for their MHCDs experiment, but compared to the gas pressure of $0.1\,\textnormal{mTorr}$
used by Alexeff and Dyer in their pen sized coaxial configuration,
this is still larger by a factor of ten thousand. Consequently, in
the aforementioned MHCDs experiments, the DC glow discharges involve
much weaker static electric fields due to larger screening effects.
When a gas filled medium is applied with an external static electric
field, the polarization process sets in and such screening effect
weakens the net electric field strength in the region filled with
gas. The degree of such screening process grows with gas pressure.
Consequently, in typical MHCDs experiments, where much higher gas
pressures are involved, relatively low self-pulsing frequencies are
observed as a result of larger screening effects for the applied static
electric field. Contrary to this, in the aforementioned pen sized
coaxial configuration investigated by Alexeff and Dyer, the externally
applied static electric field is only weakly screened due to the fact
that the device is maintained at a very low gas pressures. Consequently,
the ions in such device can oscillate at very high frequencies. 

Besides this screening effect which acts to weaken the externally
applied static electric field, the presence of higher gas pressure
increases the effective mass of an oscillating superparticle. The
superparticle concept has been  previously introduced in the discussion
of self-oscillations involving a plasma fireball. The dynamics of
plasma involves collective motions of its constituent atoms. In the
framework of this alternative theory\cite{4-Cho-1,4-Cho-2} elaborated
in this paper, the self-pulsing dynamics in both MHCDs experiments
and the pen sized coaxial\textit{ } configuration investigated by
Alexeff and Dyer involves the concept of oscillating superparticles.
The effective masses of such superparticles have an explicit dependence
on gas pressures. In general, the superparticle associated with the
higher gas pressure environment has a larger mass than the one associated
with the lower gas pressure environment. Consequently, the superparticle
in a typical MHCDs experiment has much larger mass than the one in
the pen sized coaxial configuration investigated by Alexeff and Dyer.
This also explains why self-pulsing frequencies are much lower in
MHCDs experiments compared to the device investigated by Alexeff and
Dyer.

\section{Concluding Remarks }

In summary, the mechanism behind self-sustained oscillations of an
ion in the DC glow discharges has been briefly discussed in the framework
of interaction between an ion and surface charges that it induces
at the bounding electrodes. Such alternative description provides
an elegant explanation to the formation of plasma fireballs in the
laboratory plasma. It has been found that oscillation frequencies
also increase with ion's surface charge density, but at the rate which
is significantly slower than it does with electric field strength.
Such result supports the conclusions by Fox\cite{0-Fox-1931} and
Bošan \textsl{et} \textsl{al}.,\cite{Bosan-1988} where they reported
of oscillation frequencies being not too dependent on the type of
ions used in the discharge. Self-sustained oscillations in the DC
glow discharges can be expected to have high thermal stability due
to ion's kinetic energies that are much larger than the average room
temperature thermal energies. It is well known that self-sustained
oscillations in the DC glow discharges occur only in the negative
resistance region of the voltage-current characteristic curve. Experimentally,
such region is characterized by $V_{\textnormal{th1}}\leq V_{T}\leq V_{\textnormal{th2}},$
where $V_{\textnormal{th1}}$ and $V_{\textnormal{th2}}$ are some
threshold voltages. Such observation is quite elegantly explained
from the solutions of Eq. (\ref{eq:ODE-newton}), where the physical
solutions are only found for $V_{\textnormal{th1}}\leq V_{T}\leq V_{\textnormal{th2}}.$
Presented mechanism also correctly describes the self-sustained oscillations
of ions in dusty plasmas. To demonstrate this, Eq. (\ref{eq:ODE-newton})
has been applied to correctly predict the frequency of dust particle
oscillations in the dusty plasmas experiment by Arnas \textit{et al}.\cite{dusty-plasma-Arnas - 1999}
Such result demonstrates that self-sustained oscillations in dusty
plasmas and DC glow discharges involve common physical processes.
This is the first theory to successfully explain the self-sustained
oscillations phenomena in both dusty plasmas and the DC glow corona
physics.

\section*{\noindent APPENDIX}

\subsection{\noindent Sheath potential}

\begin{figure}[h]
\begin{centering}
\includegraphics[width=0.95\columnwidth]{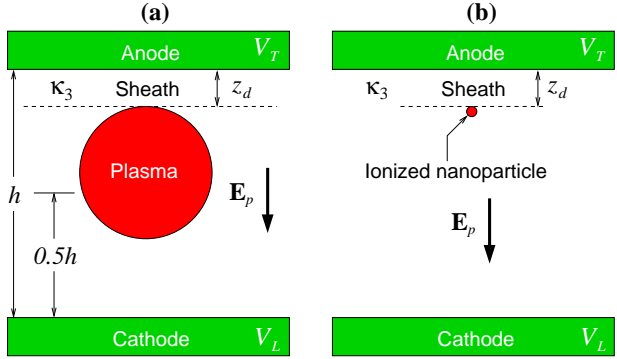}
\par\end{centering}

\caption{(Color online) (a) Illustration of the plasma and the sheath. The
plasma ball effectively behaves as one very large charged superparticle.
In the illustration, the plasma has been deliberately drawn as a sphere
to emphasize a plasma ball. In reality, however, the plasma can be
in any shape depending on the geometry of confining electrodes. (b)
The single ionized particle configuration considered in this article.
In (a) and (b), only the sheath near the anode is shown for illustration.
\label{fig:5_appendice}}
\end{figure}

It is worthwhile to compare the potential energy well of Fig. \ref{fig:2}
with the empirical potential well introduced by Tomme \textit{et al}.\cite{dusty-plasma-Tomme-sheat-potential}
 to describe the potential in the plasma sheath. The plasma sheath
is an empty space residing between the plasma and the confining electrodes.
One such plasma sheath near the anode is illustrated in Fig. \ref{fig:5_appendice}(a).
For the reason that electrons and ions move at different velocities
due to the difference in their masses, plasmas are never completely
neutral at any instant. Consequently, a plasma confined between DC
voltage biased electrodes effectively behaves as if it were a super
large charged-particle. For brevity, such ``super large charged-particle''
shall be simply referred to as a ``plasma ball'' throughout the
discussion here. In fact, the single ionized nanoparticle case illustrated
in Fig. \ref{fig:5_appendice}(b) is the limit in which the plasma
in Fig. \ref{fig:5_appendice}(a) reduces down to contain just one
ionized nanoparticle. That said, just as single ionized nanoparticle
self-oscillates when confined by DC voltage biased electrodes, the
 plasma ball also self-oscillates between the anode and cathode electrodes.\cite{1-sheat-oscillation-1996-EXP}
However, due to its large mass, the plasma ball oscillates at much
lower frequencies compared to the nanoparticle counterparts. For instance,
assuming the gap $h$ between the plate electrodes is in the order
of sub-millimeters or so in Fig. \ref{fig:5_appendice}(a), the macroscopic
plasma confined between such electrodes can contain very large number,
i.e., say, millions or more, of ionized nanoparticles or atoms depending
on the gas pressure. In general, for the study of self-sustained oscillations,
the plasma ball illustrated in Fig. \ref{fig:5_appendice}(a) can
be effectively modeled using an ionized superparticle, where the terminology
superparticle refers to a particle which is extremely large and enormously
heavier compared to the ionized nanoparticle (or atom) illustrated
in Fig. \ref{fig:5_appendice}(b). In such model, the potential in
the plasma sheath is represented by the one presented in this article.
It is quite remarkable that the sheath potential introduced empirically
by Tomme \textit{et al}. closely resembles the potential described
in this work. 

\subsection{\noindent Parameters in experiment by Arnas \textit{et al}.}

Arnas \textit{et al}.\cite{dusty-plasma-Arnas - 1999} reported that
a dust particle made of hollow glass microsphere of radius $r\approx32\,\mu\textnormal{m}$
with mass density $\rho_{m}\approx110\,\textnormal{kg}/\textnormal{m}^{3}$
and carrying a total charge of $Q\approx-4.3\times10^{5}q_{e},$ where
$q_{e}=1.602\times10^{-19}\,\textnormal{C},$ self-oscillates at frequency
of $\nu_{\textnormal{osc}}\approx17\,\textnormal{Hz}$ inside the
plasma sheath.  The inner environment of the plasma sheath, in many
respects, is similar to the environment in an empty space between
the DC voltage biased plane-parallel conductors. In that regard, the
undamped self-sustained oscillations of charged glass microsphere
in the experiment by Arnas \textit{et al}. should be theoretically
describable using the model presented in this article, where a core-shell
structured charged particle is confined between the DC voltage biased
plane-parallel conductors. The glass microsphere used by Arnas \textit{et
al}. is not core-shell structured, of course. Such dielectric particle
is obtained in the limit the radius of conductive core portion of
the core-shell structured particle goes to zero, i.e., $a=0.$ Consequently,
$\sigma_{1}$ also vanishes in that limit. That said, the glass microsphere
of radius $b=32\,\mu\textnormal{m}$ with mass density $\rho_{m}\approx110\,\textnormal{kg}/\textnormal{m}^{3}$
has total mass of $m=\left(4/3\right)\pi b^{3}\rho_{m}$ or $m\approx1.51\times10^{-11}\,\textnormal{kg}.$
Although Arnas \textit{et al}. specifically uses the word ``hollow
glass microsphere'' in their report, they do not provide any physical
details of the particle other than its outer radius. Thus, in the
aforementioned calculation of the particle's mass, I have assumed
a solid glass microsphere. The glass microsphere carries a total charge
of $Q\approx-4.3\times10^{5}q_{e},$ where $q_{e}=1.602\times10^{-19}\,\textnormal{C}.$
The surface charge density at the radius $r=b$ is hence given by
$\sigma_{2}=Q/\left(4\pi b^{2}\right)$ or $\sigma_{2}=-5.35\,\mu\textnormal{C}/\textnormal{m}^{2}.$
Arnas \textit{et al}. does not provide any information regarding the
dielectric constant $\kappa_{2}$ for their glass microsphere. Therefore,
$\kappa_{2}=6.5$ has been chosen for the dielectric constant of glass
microsphere, which value is typical of glass microspheres. For convenience,
it shall be assumed that the medium in which the glass microsphere
oscillates is a vacuum, i.e., $\kappa_{3}=1.$ The parameters in the
experiment by Arnas \textit{et al}. is summarized here for reference:
\begin{equation}
\left\{ \begin{array}{c}
\kappa_{2}=6.5,\;\;\kappa_{3}=1,\;\; h=5\,\textnormal{mm},\;\; E_{p}=5.15\,\textnormal{kV}/\textnormal{m},\\
\sigma_{1}=0\,\textnormal{C}/\textnormal{m}^{2},\;\;\sigma_{2}=-5.35\,\mu\textnormal{C}/\textnormal{m}^{2},\\
m\approx15.1\,\textnormal{ng},\;\; a=0\,\textnormal{m},\;\; b=32\,\mu\textnormal{m},
\end{array}\right.\label{eq:Arnas-parameter-1}
\end{equation}
 where the mass is in nanograms, i.e., $m\approx1.51\times10^{-11}\,\textnormal{kg}=15.1\,\textnormal{ng}.$
The initial conditions, 
\begin{equation}
z_{d}\left(0\right)=4.4\,\textnormal{mm}\quad\textnormal{and}\quad\frac{dz_{d}\left(0\right)}{dt}=0\,\frac{\textnormal{mm}}{\textnormal{s}},\label{eq:Arnas_IC}
\end{equation}
 have been chosen purely out of convenience. With Eqs. (\ref{eq:Arnas-parameter-1})
and (\ref{eq:Arnas_IC}), the equation of motion defined in Eq. (\ref{eq:ODE-newton})
can be solved numerically via Runge-Kutta method to obtain the results
illustrated in Fig. \ref{fig:4}. One can readily verify that the
theory agrees with the experiment.\cite{dusty-plasma-Arnas - 1999}

\subsection{\noindent Comparison to the potential energy of negatively charged
glass microsphere measured by Arnas \textit{et al}.}

\begin{figure}[h]
\begin{centering}
\includegraphics[width=0.9\columnwidth]{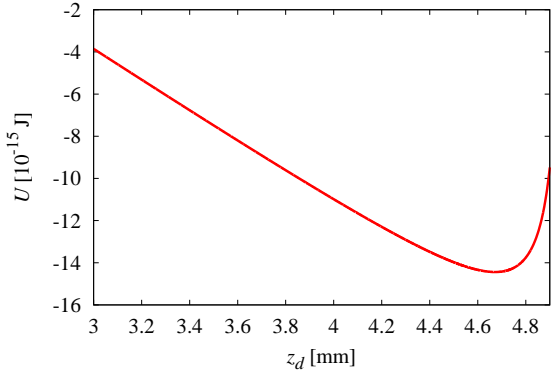}
\par\end{centering}

\caption{The potential energy of Eq. (\ref{eq:U(zd)}) is plotted for parameter
values defined in Eq. (\ref{eq:Arnas-potential-parameter-values}).
In the plot, the anode is located at $z_{d}=0\,\textnormal{mm}$ and
the cathode is located at $z_{d}=5\,\textnormal{mm}.$ This result
should be compared with the potential energy measured by Arnas \textit{et
al}.\cite{dusty-plasma-Arnas - 2000}  in their dusty plasmas experiment.
One can verify that two results are remarkably similar in potential
energy profile as well as in the order of magnitudes. \label{fig:6_appendice}}
\end{figure}

According to the model  elaborated here,\cite{4-Cho-2} the positively
charged particle confined between a DC voltage biased plane-parallel
conductors results in the formation of potential energy well in vicinity
of the anode whereas a negatively charged particle results in the
formation of potential energy well in vicinity of the cathode. Arnas
\textit{et al}.\cite{dusty-plasma-Arnas - 2000} have experimentally
verified such potential energy well for the case of negatively charged
particle in the plasma sheath near the cathode. As explained previously,
the problem of charged particle inside the plasma sheath can be effectively
modeled by a charged particle confined by the DC voltage biased plane-parallel
conductors. The potential energy function $U\left(z_{d}\right)$ of
Eq. (\ref{eq:U(zd)}) has been plotted for the following parameter
values: 
\begin{equation}
\left\{ \begin{array}{c}
\kappa_{2}=6.5,\;\;\kappa_{3}=1,\;\; h=5\,\textnormal{mm},\;\; E_{p}=3.15\,\textnormal{kV}/\textnormal{m},\\
\sigma_{1}=0\,\textnormal{C}/\textnormal{m}^{2},\;\;\sigma_{2}=-5.795\,\mu\textnormal{C}/\textnormal{m}^{2},\\
m\approx4.91\times10^{-12}\,\textnormal{kg},\;\; a=0\,\textnormal{m},\;\; b=22\,\mu\textnormal{m}.
\end{array}\right.\label{eq:Arnas-potential-parameter-values}
\end{equation}
 The result is shown in Fig. \ref{fig:6_appendice}, where it shows
the formation of potential energy well in vicinity of the cathode.
One notices that the order of magnitude for the potential energy well
is comparable to the experiment by Arnas \textit{et al}. The minor
discrepancies in the potential energy well of Fig. \ref{fig:6_appendice}
and the one measured by Arnas \textit{et al}.\cite{dusty-plasma-Arnas - 2000}
can be attributed to the fact that the electric field is not constant
in the plasma sheath whereas, between the DC voltage biased plane-parallel
conductors, electric field is a constant. 

\subsection{\noindent Physical and unphysical oscillatory solutions}

\begin{figure}[h]
\begin{centering}
\includegraphics[width=0.9\columnwidth]{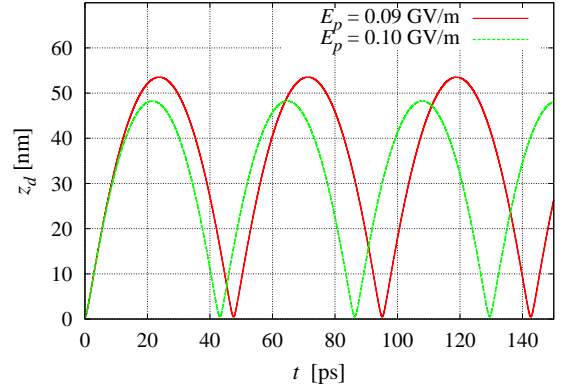}
\par\end{centering}

\caption{(Color online) Plot of $z_{d}\left(t\right)$ for $E_{p}=0.1\,\textnormal{GV}/\textnormal{m}$
and $E_{p}=0.09\,\textnormal{GV}/\textnormal{m}.$ The surface of
anode is at $z_{d}=0\,\textnormal{nm},$ the surface of cathode is
at $z_{d}=100\,\textnormal{nm},$ and the midway between the two electrodes
is at $z_{d}=50\,\textnormal{nm}.$ \label{fig:7_appendice}}
\end{figure}

The oscillatory solutions obtained for $E_{p}\lesssim0.1\,\textnormal{GV}/\textnormal{m}$
and $E_{p}\gtrsim4.9\,\textnormal{GV}/\textnormal{m}$ are unphysical
in Fig. \ref{fig:3}(d). To demonstrate this, Eq. (\ref{eq:ODE-newton})
is numerically solved and plotted for $E_{p}\lesssim0.1\,\textnormal{GV}/\textnormal{m}$
and $E_{p}\gtrsim4.9\,\textnormal{GV}/\textnormal{m}$ using the following
initial conditions and parameter values: 
\begin{equation}
\left\{ \begin{array}{c}
\kappa_{2}=\infty,\;\;\kappa_{3}=1,\;\; h=100\,\textnormal{nm},\;\; V_{L}=0\,\textnormal{V},\\
\sigma_{1}\approx3.19\,\textnormal{C}/\textnormal{m}^{2},\;\;\sigma_{2}=0\,\textnormal{C}/\textnormal{m}^{2},\\
m\approx8.95\times10^{-24}\,\textnormal{kg},\;\; b=a=1\,\textnormal{nm},\\
z_{d}\left(0\right)=0.5\,\textnormal{nm},\;\;\dot{z}_{d}\left(0\right)=0\,\textnormal{nm}/\textnormal{s},\;\;\dot{z}_{d}\equiv dz_{d}/dt.
\end{array}\right.\label{eq:IC-positive-1}
\end{equation}
 To show that $E_{p}\lesssim0.1\,\textnormal{GV}/\textnormal{m}$
yields unphysical solutions, $z_{d}\left(t\right)$ is plotted for
$E_{p}=0.1\,\textnormal{GV}/\textnormal{m}$ and $E_{p}=0.09\,\textnormal{GV}/\textnormal{m}.$
The results are shown in Fig. \ref{fig:7_appendice}, where the surface
of anode is at $z_{d}=0\,\textnormal{nm},$ the surface of cathode
is at $z_{d}=100\,\textnormal{nm},$ and the midway between the plates
is at $z_{d}=50\,\textnormal{nm}.$ It can be clearly seen that for
$E_{p}=0.09\,\textnormal{GV}/\textnormal{m},$ positively ionized
particle periodically crosses the midway between the two electrodes.
Such case is unphysical because, for positively ionized particles,
oscillations can only exist for $z_{d}<h/2-b.$\cite{4-Cho-1} Now,
it can be verified that any $E_{p}$ less than $E_{p}=0.09\,\textnormal{GV}/\textnormal{m}$
yields such unphysical solutions. 

\begin{figure}[h]
\begin{centering}
\includegraphics[width=0.9\columnwidth]{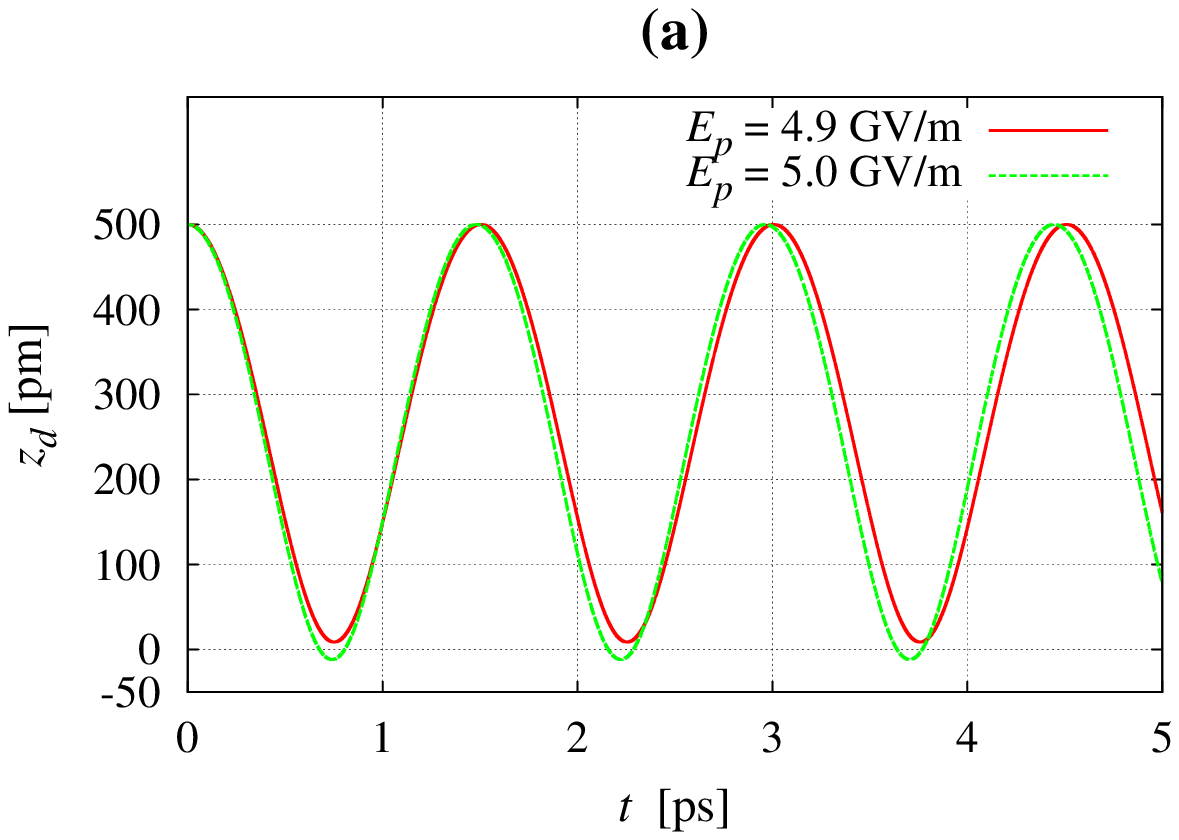}
\par\end{centering}

\begin{centering}
\includegraphics[width=0.9\columnwidth]{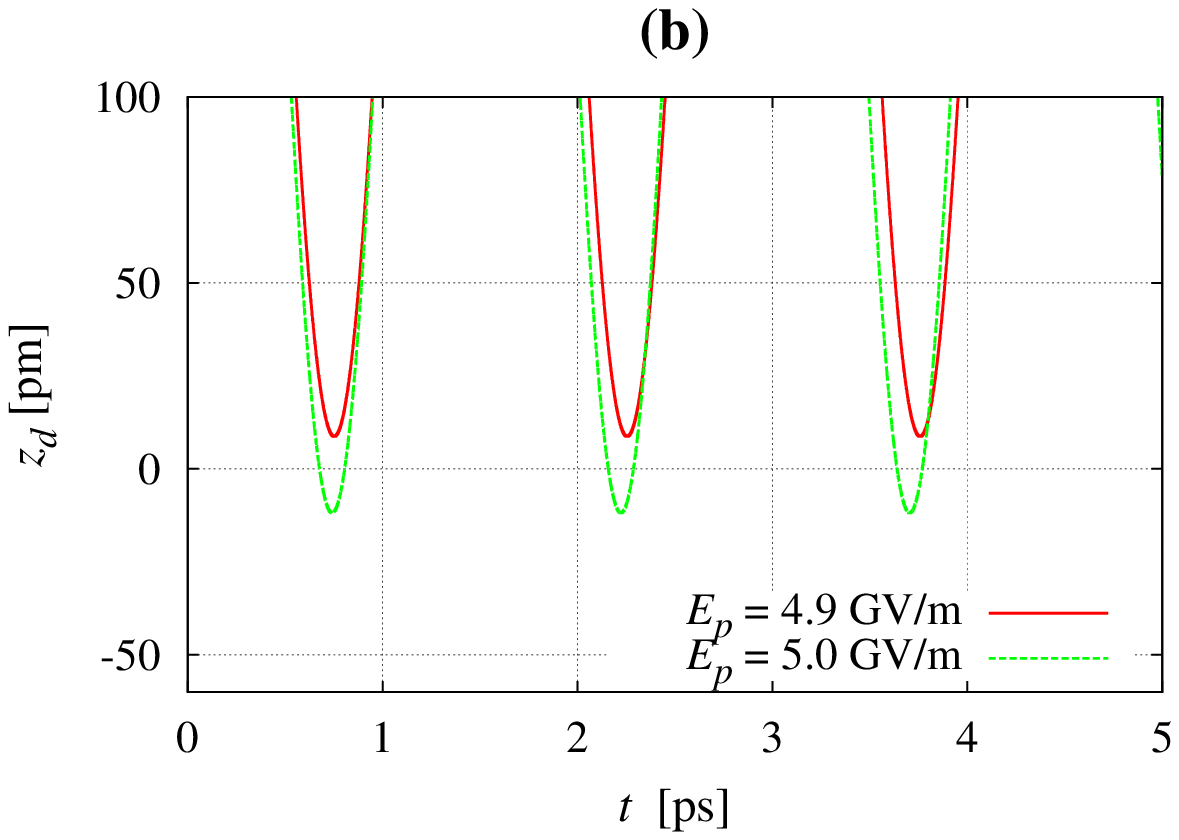}
\par\end{centering}

\caption{(Color online) (a) Plot of $z_{d}\left(t\right)$ for $E_{p}=4.9\,\textnormal{GV}/\textnormal{m}$
and $E_{p}=5.0\,\textnormal{GV}/\textnormal{m}.$ (b) The plot in
(a) has been enlarged for a view near $z_{d}=0\,\textnormal{pm}.$
The surface of anode is at $z_{d}=0\,\textnormal{pm}.$ \label{fig:8_appendice}}
\end{figure}

To show that $E_{p}\gtrsim4.9\,\textnormal{GV}/\textnormal{m}$ yields
unphysical solutions, $z_{d}\left(t\right)$ is plotted for $E_{p}=4.9\,\textnormal{GV}/\textnormal{m}$
and $E_{p}=5.0\,\textnormal{GV}/\textnormal{m}.$ The results are
shown in Fig. \ref{fig:8_appendice}(a), where the plot has been enlarged
for a view near $z_{d}=0\,\textnormal{pm}$ in Fig. \ref{fig:8_appendice}(b).
It can be clearly seen that for $E_{p}=5.0\,\textnormal{GV}/\textnormal{m},$
positively ionized particle periodically penetrates into the surface
of anode, which is unphysical. Now, it can be verified that any $E_{p}$
larger than $E_{p}=5.0\,\textnormal{GV}/\textnormal{m}$ yields such
unphysical solutions. 

For the initial conditions and parameter values defined in Eq. (\ref{eq:IC-positive-1}),
physical solutions for the oscillatory motion of positively ionized
particle are only found for $E_{p}$ satisfying the condition given
by   $0.1\,\textnormal{GV}/\textnormal{m}\lesssim E_{p}\lesssim4.9\,\textnormal{GV}/\textnormal{m},$
where the upper and lower bounds of $E_{p}$ are rough estimates based
on the grounds of the discussed physicality. This result explains
why oscillations suddenly appear at certain ``initial'' threshold
DC bias voltage and disappear suddenly when bias voltage goes beyond
certain larger ``final'' threshold voltage in experiments.

\subsection{\noindent Weakly ionized particle confined between anode and cathode
plates separated by gap of $h=1\,\textnormal{mm}$}

The oscillation frequency plot of Fig. \ref{fig:3}(d) has been obtained
for an highly ionized particle which is confined between two plate
electrodes with very small separation gap of $h=100\,\textnormal{nm}.$
Here, the same calculation is done for a weakly ionized particle confined
between plate electrodes with microscopically very large separation
gap of $h=1\,\textnormal{mm}.$ According to Chen \textit{et al}.,\cite{argon-cluster-2009}
a spherical argon cluster of radius $r=a=4.7\,\textnormal{nm}$ contains
roughly $\sim9600$ argon atoms at backing pressure of $50$ bars.
Since the particle is assumed to be weakly ionized, I shall assume
that $N=20$ electrons are removed from it. This corresponds to surface
charge density of $\sigma_{1}=Nq_{e}/\left(4\pi a^{2}\right)$ or
$\sigma_{1}\approx1.1543\times10^{-2}\,\textnormal{C}/\textnormal{m}^{2},$
where $q_{e}=1.602\times10^{-19}\,\textnormal{C}$ is the fundamental
charge unit. Neglecting the masses of missing $N$ electrons, the
particle has a total mass of $m\approx6.365\times10^{-22}\,\textnormal{kg}.$
For the calculation of particle's mass, it has been assumed that single
argon atom has mass of $6.63\times10^{-26}\,\textnormal{kg}.$ That
said, the following initial conditions and parameter values are used
for the calculation of particle's oscillation frequency as function
of electric field strength: 

\begin{equation}
\left\{ \begin{array}{c}
\kappa_{2}=\infty,\;\;\kappa_{3}=1,\;\; h=1\,\textnormal{mm},\;\; V_{L}=0\,\textnormal{V},\\
\sigma_{1}\approx11.543\,\textnormal{mC}/\textnormal{m}^{2},\;\;\sigma_{2}=0\,\textnormal{C}/\textnormal{m}^{2},\\
m\approx6.365\times10^{-22}\,\textnormal{kg},\;\; b=a=4.7\,\textnormal{nm},\\
z_{d}\left(0\right)=10\,\mu\textnormal{m},\;\;\dot{z}_{d}\left(0\right)=0\,\mu\textnormal{m}/\textnormal{s},\;\;\dot{z}_{d}\equiv dz_{d}/dt.
\end{array}\right.\label{eq:parameter-values-weak-ionization}
\end{equation}

\begin{figure}[h]
\begin{centering}
\includegraphics[width=0.9\columnwidth]{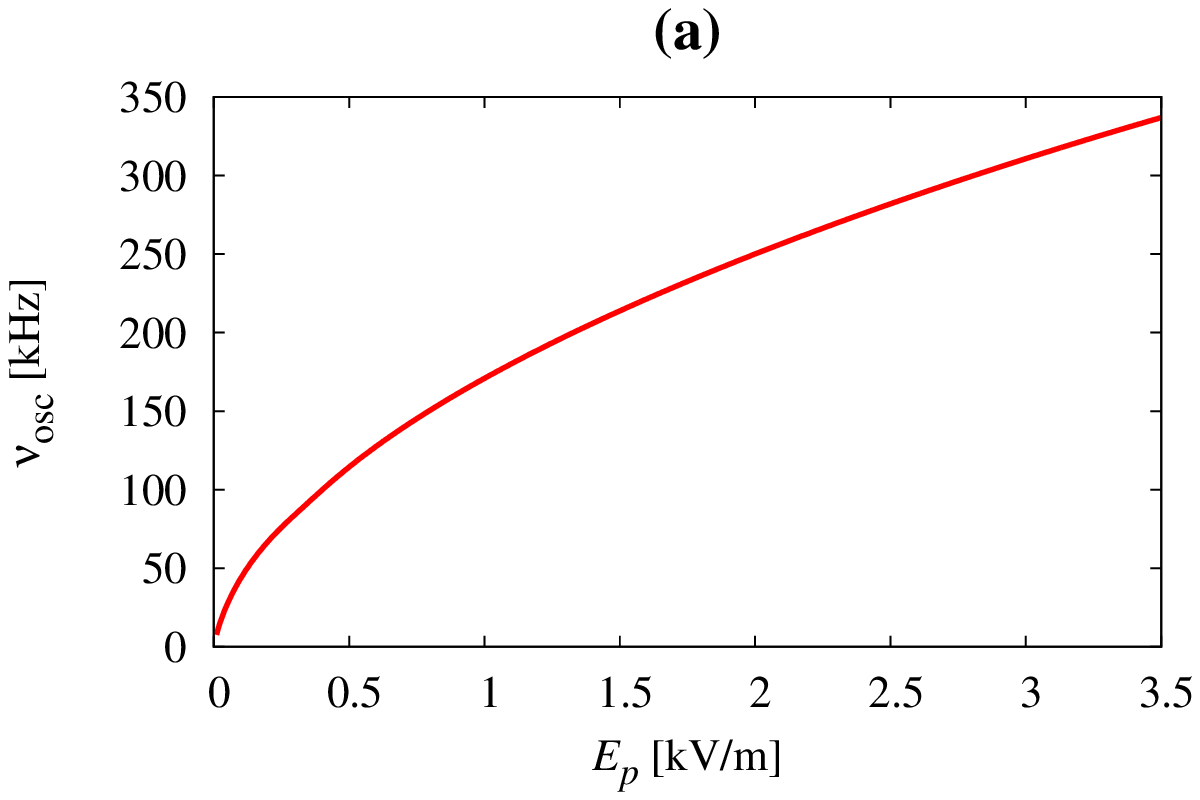}
\par\end{centering}

\begin{centering}
\includegraphics[width=0.9\columnwidth]{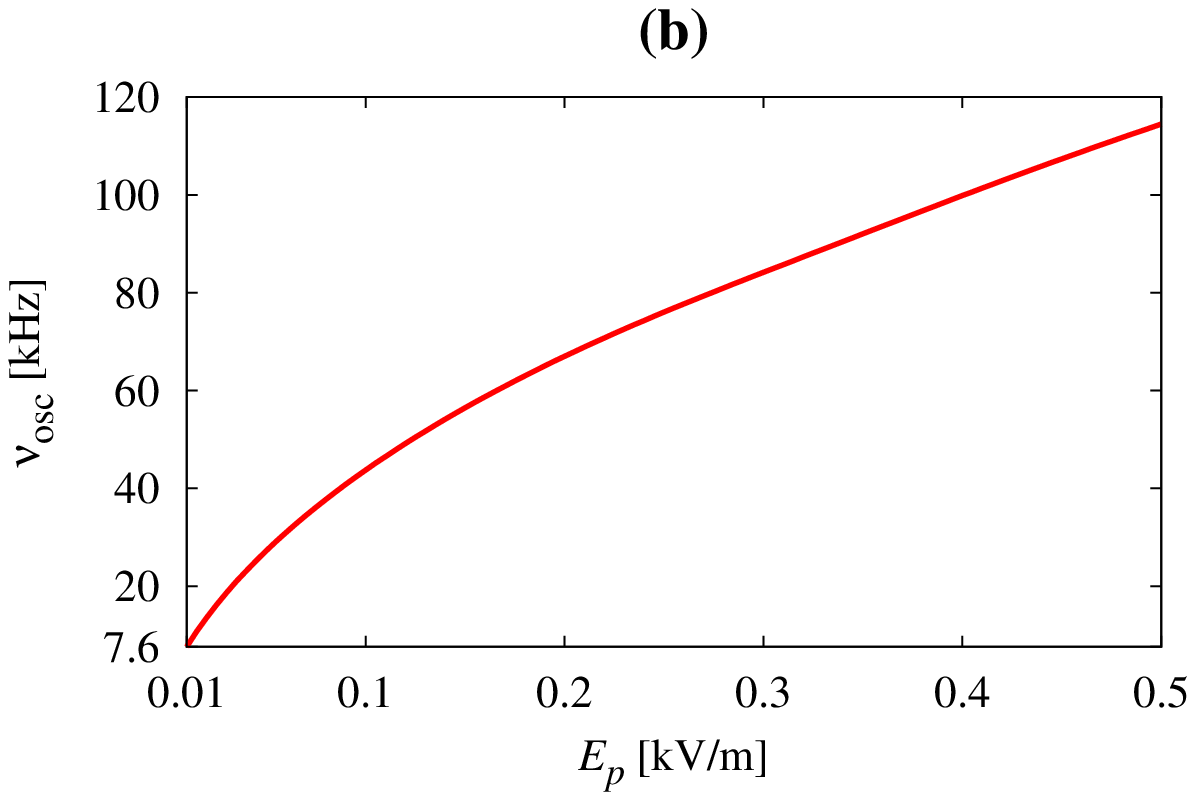}
\par\end{centering}

\caption{(a) The frequency of oscillations is plotted against electric field
strength for a weakly ionized particle with initial conditions and
parameters values defined in Eq. (\ref{eq:parameter-values-weak-ionization}).
(b) The plot in (a) is enlarged near zero. Unlike the case illustrated
in Fig. \ref{fig:3}(d), where ionized particle is confined between
electrodes with very small gap, the result here involves very large
number of discretized time steps for the Runge-Kutta routine. For
that reason, the oscillation frequency has been plotted from $E_{p}=0.01\,\textnormal{kV}/\textnormal{m}$
to $E_{p}=3.5\,\textnormal{kV}/\textnormal{m}$ in (a). The threshold
electric field strengths, i.e., $E_{p,\textnormal{th1}}$ and $E_{p,\textnormal{th2}},$
are not shown in the figure. \label{fig:9_appendice}}
\end{figure}

\noindent Using these, the equation of motion defined in Eq. (\ref{eq:ODE-newton})
is solved numerically via Runge-Kutta method for the oscillation frequency
as function of electric field strength; and, the result is shown in
Fig. \ref{fig:9_appendice}. 

This result should be compared with the one provided by Akishev \textit{et
al}.,\cite{2-Akishev-1999-FM} where they have plotted both experimental
and calculated period $T_{\textnormal{osc}}$ of self-sustained oscillations
against the average corona current at different pressures of nitrogen.
Fox\cite{0-Fox-1931} and Bošan \textsl{et} \textsl{al}.\cite{Bosan-1988}
have shown that oscillation frequencies are not very dependent on
the type of ionized particles used in the discharge. That said, the
result obtained by Akishev \textit{et al}. for nitrogen gas can be
compared with the calculation done here for ionized spherical argon
cluster. Recalling that $\nu_{\textnormal{osc}}=1/T_{\textnormal{osc}}$
and the average corona current in the electrode increases with electric
field strength, it can be readily convinced that the profile of result
shown in Fig. \ref{fig:9_appendice} is consistent with the result
obtained by Akishev \textit{et al}. 

Although the profile of oscillation frequency dependence on electric
field strength (or corona current in the electrode) is comparable
in both results, the measurement by Akishev \textit{et al}. shows
much lower oscillation frequencies for given range of electric field
strengths. Why? Such discrepancy arises from the fact that in the
experiment by Akishev \textit{et al}., the self-oscillating object
is a charged plasma ball, i.e., charged superparticle, whereas in
the calculation of Fig. \ref{fig:9_appendice}, the self-oscillating
object is a charged nanoparticle. As already discussed in section
A, a macroscopic plasma ball contains very large number of ionized
nanoparticles (or atoms) depending on the gas pressure. This makes
macroscopic plasma ball effectively a single charged superparticle
with very large mass. It has been illustrated in Fig. \ref{fig:4}
that an ionized particle with smaller mass requires weaker electric
field strength compared to the one with larger mass to oscillate at
the same frequency. Mathematically, such characteristic arises from
the fact that Eq. (\ref{eq:ODE-newton}) has mass dependence in the
denominator. Consequently, the plasma ball in the experiment by Akishev
\textit{et al}., which effectively behaves as a single charged superparticle,
oscillates at much lower frequencies for the given range of electric
field strengths compared to the configuration used in Fig. \ref{fig:9_appendice},
where the mass of an ionized particle is much smaller than the effective
mass of a plasma ball. In principle, once the mass and the effective
total charge information of the plasma ball, i.e., superparticle,
are provided, the oscillation frequency dependence on corona current
in the electrode (or electric field strength) measured by Akishev
\textit{et al}. can be reproduced by the presented theory.

\end{document}